\documentclass{article}  
\usepackage{bigsky2007}
\usepackage{graphicx}

\newcommand{\bc}{\begin{center}}
\newcommand{\ec}{\end{center}}
\newcommand{\be}{\begin{equation}}
\newcommand{\ee}{\end{equation}}
\newcommand{\bea}{\begin{eqnarray}}
\newcommand{\eea}{\end{eqnarray}}

\frompage{000} \topage{000}                                              
\def\rightmark{Perfect Fluidity}
\def\leftmark{T.~Sch\"afer}
 
\title{Perfect Fluidity in Atomic Physics} 
\authors{
{Thomas Sch\"afer %
}\\[2.812mm]
{\normalsize
Department of Physics, North Carolina State University, \\ 
Raleigh, NC 27695}}
 
\abstract{Experimental results obtained at the Relativistic
Heavy Ion Collider (RHIC) have been interpreted in terms
of a strongly interacting quark gluon plasma. The strongly
interacting plasma is characterized by ``perfect fluidity'',
i.e.~a ratio of shear viscosity to entropy density that 
saturates a proposed lower bound. In this contribution
we explore the possibility that a similar phenomenon takes
place in a strongly coupled non-relativistic Fermi liquid
in which the scattering length between the Fermions is 
infinitely large.}

 
\begin{document}
 
\maketitle
\setcounter{page}{1}
\section{Introduction}
\label{sec_intro}
  
 Experiments at the Relativistic Heavy Ion Collider (RHIC) indicate 
that a new state of matter is produced in high energy heavy ion 
collisions \cite{rhic:2005}. Much effort is currently devoted to 
characterizing the properties of this state, and to determining 
the nature of its low energy excitations. 

 Asymptotic freedom implies that the equation of state of a quark 
gluon plasma at $T\gg\Lambda_{QCD}$ is that of a free gas of quarks 
and gluons. Numerical results from lattice QCD calculations show 
that at $T\sim 2T_c$, which is relevant to the early stages of 
heavy ion collisions at RHIC, the pressure and energy density reach 
about 85\% of the free gas limit. The 15\% reduction is consistent 
with the magnitude of the first order perturbative correction. Higher 
order terms in the perturbative expansion converge very slowly, but 
this problem can be overcome using resummation techniques 
\cite{Blaizot:2003tw}. In this framework the degrees of freedom 
are dressed quasi-quarks and quasi-gluons, and these quasi-particles 
are weakly interacting.

 Transport properties of the matter created at RHIC indicate that this 
may not be correct. Experiments at RHIC suggest that the viscosity of 
the plasma is very small, and that the opacity for high energy jets is 
very large. If the plasma is composed of weakly interacting quasi-particles
then the shear viscosity can be estimated using kinetic theory. The
result is 
\be
\label{eta_kin}
\eta \simeq \frac{1}{3} n p l ,
\ee
where $n$ is the density, $p$ is the average momentum, and $l$ is the 
mean free path. In a relativistic system the number of particles is not 
conserved, and it is more natural to express the viscosity in units of 
the entropy density rather than the density. Since the entropy per particle 
(in units of $k_B$) is of order one, this does not qualitatively change the 
numerical coefficient in equ.~(\ref{eta_kin}). A good quasi-particle is 
characterized by a small ratio of the width over the excitation energy. 
This implies that the product of the mean free path times the typical 
momentum is large, and $\eta/s$ is big. This is confirmed by a weak
coupling calculation in perturbative QCD. Arnold et al.~find
\cite{Arnold:2003zc}
\be 
\frac{\eta}{s} = \frac{5.12}{g^4\log(2.42 g^{-1})},
\ee
where $g$ is the strong coupling constant. For $\alpha_s\leq 1/3$ we get 
$\eta/s>1.75$. This result should be contrasted with the values extracted 
at RHIC, which are in the range $\eta/s<0.5$ 
\cite{Teaney:2003kp,Hirano:2005wx}.

 From a theoretical point of view the RHIC results raise the question 
of how small the viscosity can get. Clearly, $\eta/s$ decreases as the
interaction becomes stronger but there are good reasons to believe that 
the shear viscosity always remains finite. In particular, it seems
reasonable to assume that the product of the mean free path and the 
typical momentum cannot be smaller than $\hbar$ \cite{Danielewicz:1984ww}.
An interesting perspective on this issue is provided by a strong coupling 
calculation performed in the large $N_c$ limit of ${\cal N}=4$ SUSY Yang 
Mills theory. The calculation is based on the duality between strongly 
coupled gauge theory and weakly coupled string theory on $AdS_5\times S_5$ 
discovered by Maldacena \cite{Maldacena:1997re}. The thermodynamics of the 
SUSY field theory was studied by Gubser et al.~\cite{Gubser:1998nz}. They 
find that the entropy density in the strong coupling limit is 3/4 of the 
free field theory result. This implies that the equation of state is not 
drastically affected by the value of the coupling. The calculation was 
extended to transport properties by Policastro et al.~\cite{Policastro:2001yc}.
These authors find that the shear viscosity to entropy density ratio of the 
strongly coupled gauge theory is $\eta/s=\hbar/(4\pi)$. This number is quite 
consistent with the values extracted from RHIC data. Kovtun et al.~studied
the behavior of $\eta/s$ in other strongly coupled field theories
with gravity duals and conjectured that the value $\hbar/(4\pi)$ 
is a universal lower bound for $\eta/s$ \cite{Kovtun:2004de,Cohen:2007qr}.  

\section{Cold atomic gases}
\label{sec_cold}

  In order to understand the relevance of these results to the 
RHIC data it is useful to study the transport properties of other 
strongly coupled fluids that are experimentally accessible. Over the 
last ten years there has been truly remarkable progress in the 
study of cold, dilute gases of fermionic atoms in which the scattering 
length $a$ of the atoms can be controlled experimentally. These 
systems can be realized in the laboratory using Feshbach resonances, 
see \cite{Regal:2005} for a review. A small negative scattering length 
corresponds to a weak attractive interaction between the atoms. This 
case is known as the BCS (Bardeen-Cooper-Schrieffer) limit. As the 
strength of the interaction increases the scattering length becomes 
larger. The scattering length diverges at the point where a bound 
state is formed. This is called the unitarity limit, because the 
scattering cross section saturates the $s$-wave unitarity bound 
$\sigma=4\pi/k^2$. On the other side of the resonance the scattering 
length is positive. In the BEC (Bose-Einstein condensation) limit the 
interaction is strongly attractive and the fermions form deeply bound 
molecules.

 The unitarity limit is of particular interest. In this limit the 
atoms form a strongly coupled quantum liquid which exhibits universal
behavior. At zero temperature the atomic liquid is characterized by 
the mass of the atoms $m$, the density $n$, the scattering length $a$,
and the effective range $r$. A dilute gas at unitarity corresponds to 
the limit in which $a^3n\to\infty$ and $r^3 n\to 0$. This implies that 
the dependence of a physical observable on $n$ and $m$ is determined 
by simple dimensional analysis, but the overall numerical constant is a
complicated, non-perturbative quantity. At finite temperature $T$ 
properties of the fluid are universal functions of the dimensionless
variable $T/T_F$, where $T_F\sim n^{2/3}/m$.

 In cold atomic gases we can reliably compute $\eta/s$ in the BCS 
limit \cite{Massignan:2004}. The ratio is temperature dependent and 
has a minimum at $T\sim T_F$. The shear viscosity is proportional to 
$1/a^2$, and $\eta/s$ is very large in the weak coupling limit. As in 
the case of QCD there are no controlled calculations in the strong 
coupling limit $a\to\infty$. It is possible, however, to reliably 
extract $\eta/s$ from experimental data on the damping of collective 
oscillations \cite{Kavoulakis:1998,Gelman:2004,Schafer:2007pr}.

\section{Collective Oscillations}
\label{sec_coll}

 In the strong coupling limit we can assume that collective modes
are approximately described by ideal fluid dynamics. The equation 
of continuity and of momentum conservation are given by 
\bea
\frac{\partial n}{\partial t} + \vec{\nabla}\cdot\left(n\vec{v}\right) 
 &=& 0 , \\
mn \frac{\partial \vec{v}}{\partial t} 
 + mn \left(\vec{v}\cdot\vec{\nabla} \right)\vec{v} &=& 
 -\vec{\nabla}P-n\vec{\nabla}V,
\eea 
where $n$ is the number density, $m$ is the mass of the atoms, $\vec{v}$ 
is the fluid velocity, $P$ is the pressure and $V$ is the external
potential. The trapping potential is harmonic
\be 
V = \frac{m}{2}\sum_i \omega_i^2 r_i^2 .
\ee
Universality implies that the equation of state is given by $P=n^{\gamma+1}
f(T/T_F)$ with $\gamma=2/3$. The compressibility at constant entropy is  
\be
\left(\frac{\partial P}{\partial n}\right)_{S} = 
 (\gamma+1)\frac{P}{n} \ .
\ee
The equilibrium distribution $n_0$ can be determined from the hydrostatic 
equation $\vec{\nabla}P_0=-n_0\vec{\nabla}V$. At $T=0$ 
\be 
\label{ThomasFermi}
n_0(\vec{r}\,) = n_0(0) \left( 1-\sum_i \frac{r_i^2}{R_i^2}
 \right)^{1/\gamma}, \hspace{0.5cm}
 R_i^2 = \frac{2\mu}{m\omega_i^2} ,
\ee
where $\mu$ is the chemical potential. In the unitarity limit the chemical 
potential is related to the Fermi energy as $\mu=\xi E_F$, where $\xi$ is 
a universal parameter. Green function Monte Carlo calculations give $\xi
\simeq 0.44$ \cite{Chang:2003}. We consider small oscillations $n=n_0+
\delta n$. From the linearized continuity and Euler equation we get 
\cite{Heiselberg:2004}
\be 
\label{lin_eu}
m\frac{\partial^2\vec{v}}{\partial t^2} = 
 -\gamma\left(\vec{\nabla}\cdot\vec{v}\right)
        \left(\vec\nabla V\right)
 -\vec{\nabla}\left(\vec{v}\cdot\vec{\nabla} V\right),
\ee
where we have dropped terms of the form $\nabla_i\nabla_j\vec{v}$
that involve higher derivatives of the velocity. This equation has
simple scaling solutions of the form $v_i=a_ix_i \exp(i\omega t)$ 
(no sum over $i$). Inserting this ansatz into equ.~(\ref{lin_eu})
we get an equation that determines the eigenfrencies $\omega$. The
experiments are performed using a trapping potential with axial 
symmetry, $\omega_1=\omega_2=\omega_0$, $\omega_3=\lambda\omega_0$.
In this case we find one solution with $\omega^2 = 2\omega_0^2$ 
and two solutions with \cite{Heiselberg:2004,Stringari:2004,Bulgac:2004}
\be
\label{w_rad}
 \omega^2 = \omega_0^2\left\{ \gamma+1+\frac{\gamma+2}{2}\lambda^2 
  \pm \sqrt{ \frac{(\gamma+2)^2}{4}\lambda^4
           + (\gamma^2-3\gamma-2)\lambda^2 
           + (\gamma+1)^2 }\right\}. \nonumber 
\ee
In the unitarity limit ($\gamma=2/3$) and for a very asymmetric trap, 
$\lambda\to 0$, the eigenfrequencies are $\omega^2=2\omega_0^2$ and 
$\omega^2=(10/3)\omega_0^2$. The mode $\omega^2=(10/3)\omega_0^2$ is 
a radial breathing mode with $\vec{a} = (a,a,0)$ and the mode 
$\omega^2=2\omega_0^2$ corresponds to a radial dipole $\vec{a} = 
(a,-a,0)$.

 The frequency of the radial breathing mode agrees very well with 
experimental results \cite{Kinast:2004}. In a hydrodynamic description
the damping of collective modes is due to viscous effects. The dissipated 
energy is given by
\be
\dot{E} = -\frac{1}{2} \int d^3x\, \eta(x)\, 
  \left(\partial_iv_j+\partial_jv_i-\frac{2}{3}\delta_{ij}
      \partial_k v_k \right)^2  
   -\int d^3x \, \zeta(x)\, \big( \partial_iv_i\big)^2 
  \nonumber , 
\ee
where $\eta$ is the shear viscosity and $\zeta$ is the bulk viscosity.
In the unitarity limit the system is scale invariant and the bulk 
viscosity in the normal phase vanishes. For the radial scaling 
solutions we get
\be 
\overline{\dot{E}} = -\frac{2}{3}
  \left( a_x^2+a_y^2-a_xa_y \right) \int d^3x\, \eta(x),
\ee
where $\overline{E}$ is a time average. The damping rate is given by 
the ratio of the energy dissipated to the total energy of the collective 
mode. The kinetic energy is 
\be 
 E_{kin} = \frac{m}{2}\, \int d^3x\, n(x) \vec{v}^{\,2} 
 = \frac{mN}{2} \left( a_x^2+a_y^2 \right) 
  \langle x^2 \rangle. 
\ee
At $T=0$ we find $\langle x^2 \rangle = R_\perp^2/8$, where $R_\perp$ 
is the transverse size of the cloud. At non-zero temperature we 
can use the Virial theorem \cite{Thomas:2005} to relate $\langle x^2 
\rangle$ to the total energy of the equilibrium state, $\langle x^2
\rangle/\langle x^2\rangle_{T=0}=E/E_{T=0}$. The damping rate is 
\cite{Kavoulakis:1998,Gelman:2004}
\be 
\label{edot_e}
-\frac{1}{2} \frac{\overline{\dot{E}}}{E}   = 
 \frac{2}{3}\, \frac{a_x^2+a_y^2-a_xa_y}{a_x^2+a_y^2}\,
 \frac{\int d^3x\, \eta(x)}{mN \langle x^2\rangle} . 
\ee
Note that the second factor on the RHS is $1/2$ for the radial breathing 
mode and $3/2$ for the radial dipole mode. If this dependence could be 
demonstrated experimentally, it would confirm that the damping is indeed 
dominated by shear stress. 

\begin{figure}
\begin{center}
\includegraphics[width=8cm]{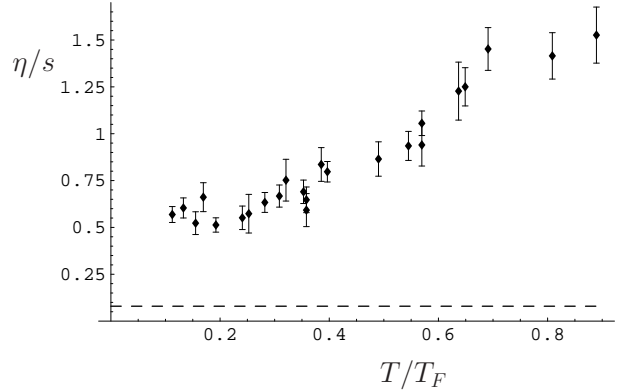}
\end{center}
\caption{\label{fig_eta_s}
Viscosity to entropy density ratio of a cold atomic gas in the unitarity 
limit, from \cite{Schafer:2007pr}. This plot is based on the damping data 
published in \cite{Kinast:2005} and the thermodynamic data in 
\cite{Kinast:2005b,Luo:2006}. The dashed line shows the conjectured 
viscosity bound $\eta/s=1/(4\pi)$. }
\end{figure}

 We shall assume that the shear viscosity is proportional to the entropy 
density, $\eta(x)=\alpha s(x)$. We note that since the flow profile has 
a simple scaling form the damping rate is proportional to the volume 
integral of the shear viscosity. If $\eta\sim s$  then the damping rate 
is proportional to the total entropy. The kinetic energy, on the other 
hand, scales with the number of particles. We can now relate the 
dimensionless damping rate $\Gamma/\omega_\perp=1/(\tau\omega_\perp)$
of the radial breathing mode to the shear viscosity to entropy density 
ratio. We find 
\be 
\label{eta_s}
\frac{\eta}{s}  =\frac{3}{4} \xi^{1/2} (3N)^{1/3} 
 \left(\frac{\bar\omega\Gamma}{\omega_\perp^2}\right)
 \left(\frac{E}{E_{T=0}}\right)
 \left(\frac{N}{S}\right).
\ee
Fig.~\ref{fig_eta_s} shows $\eta/s$ extracted from the experimental
results of the Duke group \cite{Kinast:2005}. The entropy per particle
was also taken from experiment \cite{Luo:2006}. Similar results are
obtained if the entropy is extracted from quantum Monte Carlo data. 
The critical temperature for the superfluid/normal transition is 
$T_c/T_F\simeq 0.29$. We observe that $\eta/s$ in this regime is 
roughly 1/2. This value is compatible with the conjectured viscosity
bound and comparable to the values that have been extracted at RHIC.

\section{Elliptic Flow}
\label{sec_eflow}

  Collective modes are very useful because it is possible to track
many compression and expansion cycles and even small damping coefficients 
can be measured. In heavy ion collisions we do not have this luxury and 
we have to rely on collective flow measurements to extract transport 
coefficients. In the following we shall estimate the effect of a non-zero 
shear viscosity on the elliptic flow of a cold atomic liquid. We consider 
the expansion of the atomic cloud after the trapping potential is removed. 
The expansion is described by a simple scaling ansatz
\be 
n(r_i,t) = n_0(r_i/b_i(t)) \;\;(i=1,\ldots,3), \hspace{1cm}
v_i(\vec{r},t) = \frac{\dot b_i(t)r_i}{b_i(t)}\, .
\ee
It is easy to check that this ansatz satisfies the continuity 
equation. The Euler equation gives \cite{Menotti:2002}
\be 
 \ddot b_i = \frac{\omega_i^2}{(b_1b_2b_3)^\gamma}\frac{1}{b_i}.
\ee
We are interested in an axially symmetric trap with $\omega_x=
\omega_y=\omega_\perp$ and $\omega_z=\lambda\omega_\perp$. In
the limit $b_\perp\gg b_z$ we get
\be 
\ddot b_\perp = \frac{\omega_\perp^2}{b_\perp^{1+2\gamma}},
\hspace{0.5cm}
\ddot b_z = \frac{\omega_z^2}{b_\perp^{2\gamma}}.
\ee
\begin{figure}
\begin{center}
\includegraphics[width=6cm]{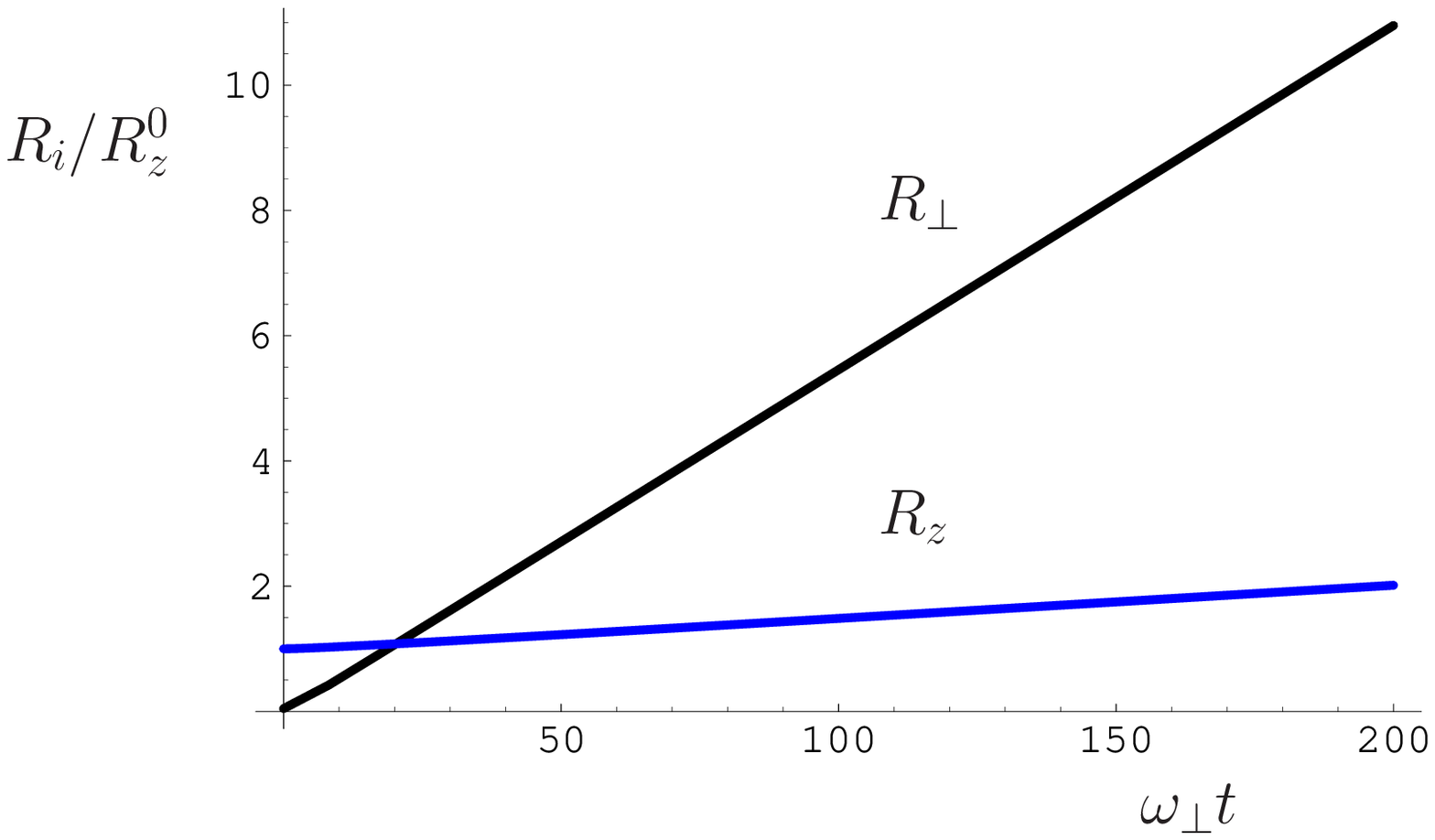}
\includegraphics[width=6cm]{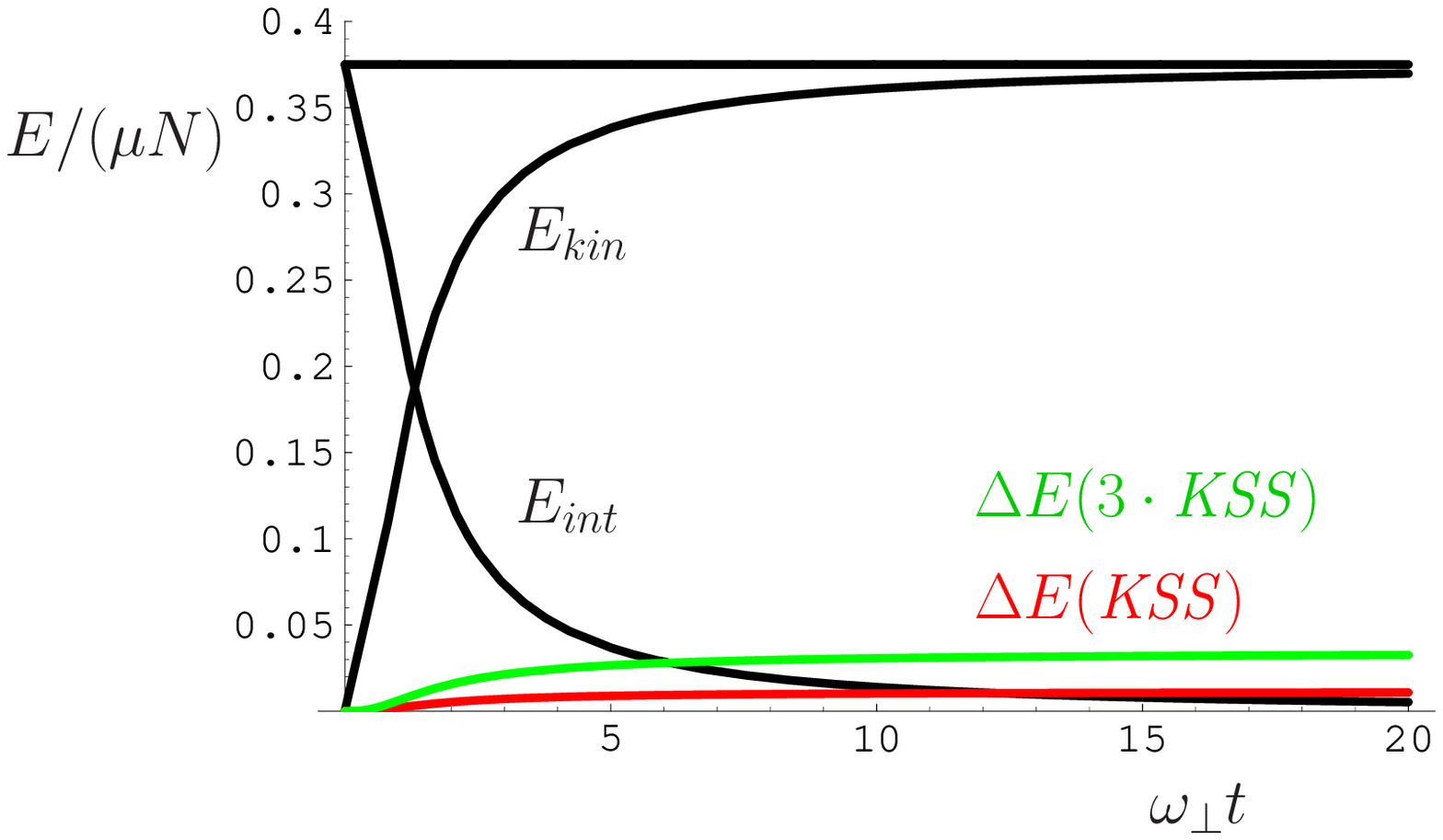}
\end{center}
\caption{\label{fig_flow}
The left panel shows the evolution of the scale factors $b_\perp$ 
and $b_z$ as a function of the dimensionless variable $\omega_\perp t$.
The right panel shows the time evolution of the internal and kinetic 
energies. We also show an estimated of the energy dissipated by 
viscous effects. The two curves correspond to $\eta/s$ equal to
one and three times the conjectured viscosity bound (using $s$ at 
$T=T_c$).}
\end{figure}
The solution of these equations for a very asymmetric trap ($\lambda
=0.045$) is shown in Fig.~\ref{fig_flow}. We observe the usual elliptic 
flow phenomenon: the transverse pressure exceeds the longitudinal 
pressure, there is more acceleration in the transverse direction, 
and as a consequence the transverse expansion is faster than the 
longitudinal one. The right panel shows that during the expansion
internal energy is converted to kinetic energy of the flow. We 
can also compute the amount of energy dissipated due to viscous 
effects. We find
\be
\dot E = -\frac{4}{3} 
\left(\frac{\dot b_\perp}{b_\perp}-\frac{\dot b_z}{b_z}\right)^2
  \ \int d^3x \ \eta(x).
\ee
An estimate of this quantity is also shown in the right panel of 
Fig.~\ref{fig_flow}. We show two curves, corresponding to $\eta/s$ 
equal to one and three times the conjectured viscosity bound, 
respectively. We have taken the entropy $s$ at $T=T_c$. We observe 
that most of the energy is dissipated early during the expansion. 
The total energy dissipated amounts to a few percent of the total 
energy available. This should make the effect observable, although it 
seems unlikely that a measurement of $\eta/s$ based on elliptic 
flow can be as accurate as the one based on collective modes. A
measurement of elliptic flow was reported in \cite{oHara:2002},
but there is no systematic study of the temperature dependence
of the effect.

Acknowledgments: This work is supported in part by the 
US Department of Energy grant DE-FG02-03ER41260.


\vfill\eject
\end{document}